%% file: nss_main.tex
\documentclass[runningheads]{llncs}
\usepackage{graphicx}
\usepackage{geometry}
\usepackage{hyperref}
 \usepackage{tikz}
\newcommand{\ballnumber}[1]{\tikz[baseline=(myanchor.base)] \node[circle,fill=.,inner sep=1pt] (myanchor) {\color{-.}\bfseries\footnotesize #1};}

\newcommand{\subheading}[1]{\noindent{\textbf{#1}}}
\begin{document}
\title{RAIDER: Reinforcement-aided  Spear Phishing Detector}
%
%
\author{Keelan Evans\inst{1,3} \and
Alsharif Abuadbba\inst{2,3} \and
Tingmin Wu\inst{2} \and 
Kristen Moore\inst{2} \and 
Mohiuddin Ahmed\inst{1} \and
Ganna Pogrebna\inst{4} \and
Surya Nepal\inst{2,3} \and
Mike Johnstone\inst{1}
}
\authorrunning{Abuadbba et al.}
\institute{Edith Cowan University, Australia \and
CSIRO's Data61, Australia\\\and
Cybersecurity Cooperative Research Centre, Australia\\\and
Charles Sturt University \& The University of Sydney \& The Alan Turing Institute\\
\email{sharif.abuadbba@data61.csiro.au}}
\maketitle              
\input{Sections/0_Abstract}


\input{Sections/1_Introduction}
\input{Sections/2_Background}

\input{Sections/3_Key_insights}
\input{Sections/4_system_Design}

\input{Sections/5_Evaluation}
\input{Sections/6_Discussion}

\input{Sections/7_Related_Work}

\input{Sections/8_Conclusion}

\bibliographystyle{unsrt}
\bibliography{references}

\end{document}

%% file: Sections/0_Abstract.tex
\graphicspath{ {./Figures/} }

\begin{abstract}

Spear Phishing is one of the most difficult to detect cyber attacks facing businesses and individuals worldwide. In recent years, considerable research has been conducted into the use of Machine Learning (ML) techniques for spear-phishing detection. ML-based solutions are vulnerable to zero-day attacks, as when the algorithms do not have access to the relevant historical data, they cannot be reliably trained. Furthermore, email address spoofing is a low-effort yet widely applied forgery technique in spear phishing which the standard email protocol SMTP fails to detect without the use of extensions. Detecting this type of spear threat requires (i) a close investigation of each sender within the mailbox; and (ii) a thorough exploration of the similarity of its characteristics to the spoofed email. This raises scalability challenges due to the growing number of features relevant for investigation and comparison, which is \textit{proportional to the number of the senders within a particular mailbox}. This differs from traditional phishing attacks, which typically look at email bodies and are generally limited to a binary classification between `\textit{phishing}' and `\textit{benign}' emails. 

We offer a possible solution to these problems, which we label RAIDER:  \underline{R}einforcement \underline{AI}ded Spear Phishing \underline{DE}tecto\underline{R}. A reinforcement-learning based feature evaluation system that can automatically find the optimum features for detecting different types of attacks. By leveraging a reward and penalty system, RAIDER allows for autonomous features selection. RAIDER also keeps the number of features to a minimum by selecting only the significant features to represent phishing emails and detect spear phishing attacks.
After extensive evaluation of RAIDER on over 11,000 emails and across 3 attack scenarios, our results suggest that using reinforcement learning to automatically identify the significant features could reduce the dimensions of the required features by $55\%$ in comparison to existing ML-based systems. It also increases the accuracy of detecting spoofing attacks by $4\%$, from $90\%$ to $94\%$. Furthermore, RAIDER demonstrates reasonable detection accuracy against  a sophisticated attack named ``Known Sender'', in which spear phishing emails greatly resemble those of the impersonated sender. By evaluating and updating the feature set, RAIDER is able to increase accuracy by close to $15\%$, from $49\%$ to $62\%$  when detecting Known Sender attacks. 

\end{abstract}

%% file: Sections/1_Introduction.tex
\section{Introduction}\label{sec:intro} 

Phishing is a type of cyber attack in which the adversary uses social engineering techniques to either convince a user to do something they should not do or motivate them to abstain from doing something they are supposed to do. In other words, adversaries attempt to disguise themselves as trusted individuals to elicit sensitive information from a target, or get them to perform some specific task like installing malware on their computer or transferring money to the attacker. Spear Phishing, a variant of phishing attacks, targets a specific individual with an attack crafted based on prior information about the target and their relationship with the impersonated sender~\cite{benenson2017unpacking}. This differs from traditional phishing attacks which blindly target a large number of people with a generic attack, and results in a more effective disguise, making it difficult in a lot of cases to discern between what is a spear phishing attack and what is a legitimate email. It is estimated that over 80\% of organisations in the US alone have experienced a spear phishing attack, and that they account for billions of dollars in losses annually~\cite{thomas2018individual}. While spear phishing attacks are effective on their own, they may also be used to gain a foothold into a network as a part of a more extensive attack \cite{internet_sec,ho2019detecting}. Spear phishing attacks are conducted across many forms of Internet communication but are most commonly delivered by means of an email \cite{kim2020deepcapture}.

There are a variety of different spear phishing techniques used by attackers to masquerade themselves as another (trusted) sender. Effective spear phishing attacks are two-pronged attacks which involve both \textbf{(1)} manipulating the headers of an email to forge certain fields to more closely resemble an email of the sender they are trying to impersonate, and \textbf{(2)} psychologically manipulating the recipient of the email (e.g., by mentioning a third party or an individual a potential victim trusts, enticing a sense of urgency, using common knowledge facts, etc.). Email address spoofing is one of the most common forgery techniques, requiring low effort on the part of an adversary. This technique involves manipulating the header of an email to make the email appear to be from a different sender than the individual or entity who actually sent the email. Email address spoofing in itself is not a spear phishing attack, but it is a commonly used tool in spear phishing helping the attacker to masquerade as another sender they are impersonating. Considering that email address spoofing techniques are commonly used in spear phishing emails \cite{shen2021weak}, observing spoofing could be considered a reliable indicator of a spear phishing attack in progress.      

Existing defences against spear phishing include both (a) ``patching with people'' techniques - i.e., educating end-users to identify spear phishing attacks; as well as (b) ``patching with tech'' (which is a focus of this paper) - i.e., building detection algorithms for reliable spear phishing identification \cite{caputo2013going,canova2014nophish} and development of software solutions to scan emails and predict whether they are spear phishing or benign. Software solutions can make use of blacklists of IP addresses and URLs associated with phishing attacks \cite{ghafir2015advanced,ramachandran2007filtering,parmar2012protecting}, as well as Machine-Learning (ML) based solutions \cite{gascon2018reading,dewan2014analyzing,duman2016emailprofiler,samad2020analyzing}. 

While general phishing and spear phishing are somewhat related, the design of ML based detection models for them differs significantly. Unlike general phishing where ML detection is a binary classification problem with emails either being `\textit{phishing}' or `\textit{benign}' \cite{thapa2020fedemail},  spear phishing ML detection is a more complex multi-class classification problem in which each class corresponds to a sender. To create the spear phishing detection model, one, therefore, needs to look at the metadata and extract features from each class (sender) within every mailbox. Then, once a new email is received, features would be extracted from that email's metadata and the model would measure how close those features align to all possible senders to detect discrepancies between the features of the email and the features of the alleged sender. The existence of these discrepancies is an indication of a potential spear attack. 

So while on the surface the problem may seem like only a 2-class binary classification problem in which an email is being classified as either `\textit{spear phishing}' or `\textit{benign}', in actuality, there are two phases to the classification of spear phishing. The first phase is a multi-class classification problem where an incoming email is being classified as one of the existing sender classes, and the second phase is to determine whether an email is spear phishing or not based on the result of the first classification problem. Therefore the spear phishing problem is a lot more complex than the binary classification problem of general phishing attacks, with the complexity of the multi-class classification problem being variable and dependent on the number of senders within a mailbox/organisation/etc.

Despite the efficacy of ML as a defence against spear phishing, we have identified several challenges that limit the practical implementation of these solutions. Existing implementations produce large feature vectors for each individual sender within a mailbox \cite{gascon2018reading}. This limits scalability, as each new email in a mailbox will produce another large vector, which consumes significant amounts of memory and is impractical beyond small-scale implementations; that is \textit{as the number of senders increases, so does the number of classes and dimensions of the feature vectors}. Current research also neglects changes within the dataset. Zero day attacks can cause a performance reduction in classifiers as they can differ significantly from the initial training data. To the best of our knowledge, no prior research has attempted to evaluate nor update features in response to emerging threats within spear phishing. 

This motivates us to address the following research question:

\textbf{\textit{How can a spear phishing detection system operate efficiently as well as effectively over time while keeping feature vector dimensions low and continuously detecting new attacks? 
}} 

To answer this question, we explored the possibility of automated feature engineering using Reinforcement Learning (RL) ~\cite{sutton2018reinforcement}. RL is the training of ML models to make an autonomous optimum sequence of decisions in uncertain situations using rewards and penalty strategies. RL has been used in various domains such as games \cite{machado2018revisiting}, car racing simulators \cite{chou2017improving}, robotic simulator \cite{lowrey2018reinforcement}, and virtual reality \cite{zhu2017target}. In this work, we aim to leverage the autonomous RL ability  to select the optimum number of features required to achieve a reasonable balance between accuracy and low feature dimensions to detect spear phishing.

Our key contributions are summarized as follows:
\begin{itemize}
\item \textbf{(Existing Limitations)} We explore previous spear phishing classifiers and identify the key challenges facing their practical implementation. After exploring with  $32,000$ reported phishing emails over 5 years (2016-2020), we find that the dimension of feature vectors is considerably high ($>8000 $) dimensions when features are extracted manually using existing techniques \cite{gascon2018reading}, and that the importance of features and their contribution to the efficacy of the classifier is obscured. We identify that spear phishing attacks are intentionally modified over time in order to avoid detection, and that because of the nature of static learning, classifiers misinterpret previously unseen attacks.  
\item \textbf{(RAIDER)} We develop a novel RL-based feature selection model, called RAIDER\footnote{Code and artifacts will be released upon publication or reviewers request.}. To the best of our knowledge, RAIDER is the first spear phishing detector that can account for changing data over time by generating new feature subsets for specific types of attacks. This is achieved by employing a reward/penalty system that determines how much influence a feature has over prediction accuracy. RAIDER automatically chooses only the key features that allow it to best predict a variety of different spear phishing attacks.
\item \textbf{(Lightweight and Effective)} We perform an extensive evaluation of RAIDER over 11,000 emails across 3 attack scenarios: Blind spoofing, Known domain, and Known sender. Our results suggest that using RL to automatically identify the significant features could reduce the required feature dimensions by $55\%$ compared to manual feature extraction, while eliminating the time and effort required for manual feature extraction. It also allows us to achieve an accuracy improvement in spoofing attack detection by $4\%$ (an increase from $90\%$ to $94\%$).

\end{itemize}

%% file: Sections/2_Background.tex
\section{Background}
In this section, we provide the prior knowledge for K Nearest Neighbours and Reinforcement Learning.

\subsection{K Nearest Neighbours (KNN)}\label{sec:machinelearning}

\begin{figure}[h!]
\centering
\includegraphics[scale=0.5,bb=0 0 200 200]{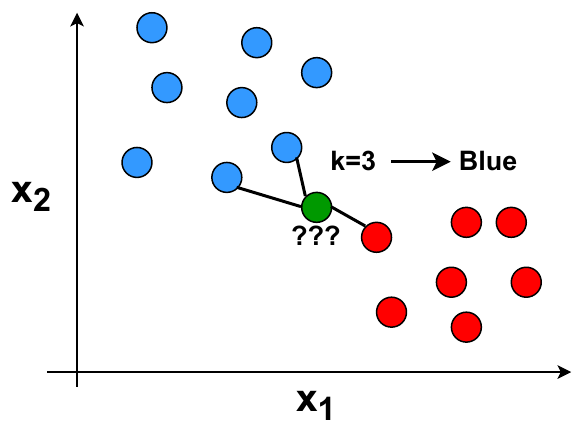}
\caption{KNN makes predictions based on a vote between the k nearest data points.} 
\label{fig:knn}
\end{figure}

KNN is a classification algorithm that groups training samples into classes and then predicts what class the incoming data belongs to \cite{cunningham2020k}. To find which class a new data point belongs to, KNN finds the $k$ most similar training samples to that data point and then facilitates a `voting' selection procedure to determine the predicted class of the data. 
An example is shown in Figure \ref{fig:knn}, $k=3$ means that the 3 data points nearest to the incoming data will be counted in the vote. Each data point `votes' on behalf of its class. 
Two of the three data points belong to the `\textit{spear phishing}' class (in blue), and the rest represents `\textit{benign}' (in red). The new data point (in green) will be classified as `\textit{spear phishing}'. The value of $k$ and the specific algorithm used to calculate the distance between points is dependent on the type of data that KNN is working with, and therefore varies between use cases.

RAIDER uses KNN to predict whether an email is spear phishing or not. It is used to determine the effect each feature has on accuracy during the RL phase, as well as to make predictions using the automatically generated feature subset. KNN has been chosen because it calculates the distances between data points when predictions are being made, not during the training phase. So while adding new data requires the KNN model to be retrained from scratch, the distances between data points do not need to be recalculated, since this is all done when making predictions.   
Specific uses of KNN are explained in more detail in Section \ref{RAID_detail}.

\subsection{Reinforcement Learning (RL)}\label{sec:reinforcementlearning}

RL is a subsection of ML that uses a penalty and reward system to train an agent to perform specific tasks in response to certain circumstances \cite{sutton2018reinforcement}. The reinforcement agent is given an environment in which it can perform actions and learn their impact. The specific situation the agent is in within the environment is called a state. The agent selects an action based on its interpretation of the state it is in at that time. Rewards and penalties are given to the agent based on the action it performs. By associating the combination of state and action with the resulting reward/penalty, the reinforcement agent learns what actions are the best to perform in each scenario. The mapping of states to particular actions is called a policy, and it defines how a reinforcement agent acts in any given situation.

For RAIDER, we propose an RL algorithm based on Fard et al.'s Average of Rewards (AOR) policy from their RL-based feature selection model~\cite{fard2013using}. In particular, they propose an approach to select the best subset of features to use in a classification model. We adapt their approach to select the best feature collection to detect email spear phishing. Each feature has an AOR, which is defined to be the average increase/decrease in accuracy incurred by using that feature. We choose the features based on exploring the action space and its impact on the bottom line accuracy by randomly selecting a new subset of features at a time. When a feature is chosen 2 times within subsets, the increase/decrease in accuracy for each of these actions will be summed and divided by 2. The formula for AOR is as follows:

\begin{equation}
    (AOR_f)_{New}=\left [  (k-1)(AOR_f)_{Old} + V(F)\right ]/k
\label{eq:AOR}
\end{equation}

Where $f$ is the selected feature, $(AOR_f)_{Old}$ is the current feature value in the table and $(AOR_f)_{New}$ is the updated value. $V(F)$ is the assessment value of the current state $F$ and $k$ is the number of times the feature $f$ has been used during RL training.

While, to the best of our knowledge, there have been no previous applications of RL to the detection of spear phishing emails, RL has been successfully applied to the detection of generic phishing emails. Smadi et al. \cite{smadi2018detection} used RL to dynamically update a neural network-based general phishing detection system over time in response to changes in the environment. The detection of generic phishing attacks is a binary classification problem, whereas spear phishing is commonly implemented as a large multi-class classification problem in which each class corresponds to a sender. Therefore, re-applying binary classification from generic phishing detection, such as in \cite{smadi2018detection}, is not sufficient for detecting spear phishing attacks. Also, due to the growing size of classes (senders), auto-learning neural network approaches are not suitable. Hence, a new RL design is required for auto-feature selection in the multi-class classification spear phishing problem.

%% file: Sections/3_Key_insights.tex
\section{Key Insights: Analysis of Spear Phishing Emails}

We build on the previous work by Gascon et al. \cite{gascon2018reading} as their research offers very promising  results for ML-based detection of spear phishing emails (these results appear to outperform many competitive models). We first reproduce Gascon et al. results by implementing a KNN-based system as reported in their paper, making use of a subset of their 46 features. Our experimental approach also utilizes the attack methods Gascon et al. used to test and evaluate their security ecosystem. By reproducing this state-of-the-art spear phishing detection mechanism, we are able to gain two insights into the challenges that are faced by traditional feature engineering in the detection of spear phishing emails: the first insight is related to the feature vector stability and the second insight to the feature importance.

\begin{figure}[h!]
\centering 
\includegraphics[width=0.5\linewidth,scale=1]{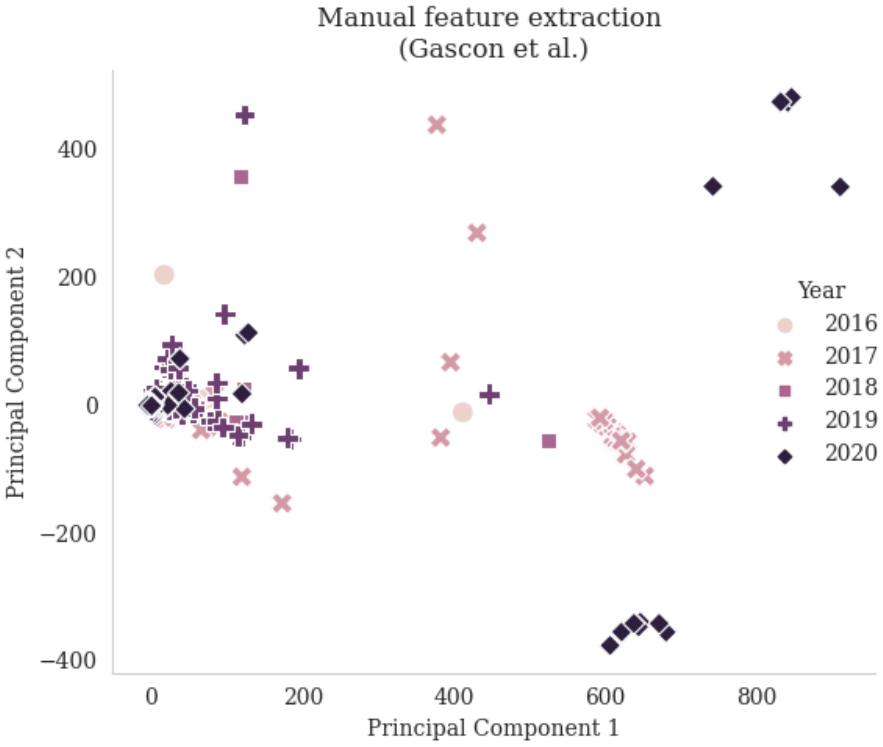}
\caption{ The visualisation of traditional feature extraction of phishing emails as in \cite{gascon2018reading} over five years (2016-2020) using the PCA (Principal Component Analysis).
Here PCA projects the data points into a new 2D space while the axes do not have physical meaning. The new values on the x-axis and y-axis (Principal Component 1 and 2) contribute most to the variation through a transformation.} 
\label{fig:gascon_PCA}
\end{figure}

\subheading{\underline{Insight (1)}: Feature Vector Stability.}
We found that the features produced by manual feature engineering are unstable, sparse and high-dimensional. This causes high memory consumption and is not a scalable solution. As more classes are added to the initial classifier (through the addition of new senders' characteristics to the model), the number of feature vectors will also grow. This can quickly become impractical and difficult to manage as there will be a significant number of high dimensional feature vectors. Figure \ref{fig:gascon_PCA} visualises the features generated by the state-of-the-art manual feature engineering over time in a 2D space using the PCA (Principal Component Analysis). The distribution of the data points illustrates that the feature vectors have different spreads, which are unstable and sparse over the years.

\subheading{\underline{Insight (2)}: Feature Importance.}
Current uses of traditional feature engineering within the field of spear phishing detection do not allow for the understanding of the importance of individual features. Manually extracted features also cannot be adjusted or improved over time without manual intervention. This is a significant problem when considering previously unseen attacks not present in the classifier's training data. It has been demonstrated that spear phishing attacks change over time, and that phishing campaigns are modified throughout their life span in order to evade detection. Heijden et al. \cite{van2019cognitive} recently demonstrated  how over the period of a phishing campaign, there were intentional attempts to modify and alter spear phishing emails in order to avoid detection. It is therefore important to have a method for autonomously evaluating the efficacy of features for detecting spear phishing attacks, and updating these features in response to changes within the data and the emergence of new threats.

This problem presents the question of \textit{how can we determine the most important features for detecting spear phishing emails while ensuring the efficiency and practicality of our solution by reducing the size of our feature vectors as much as possible?}

%% file: Sections/4_system_Design.tex
\section{RAIDER System Design}
In  this  section,  we  provide  the RAIDER model design. We first define the threat models that we focused on in this paper. We  then  provide an  overview  of  the proposed RAIDER detection  system followed by an in-depth explanation. 
\subsection{Threat Model}\label{tmodel}
\begin{figure}[h]
\centering
\includegraphics[width=1\linewidth,scale=1]{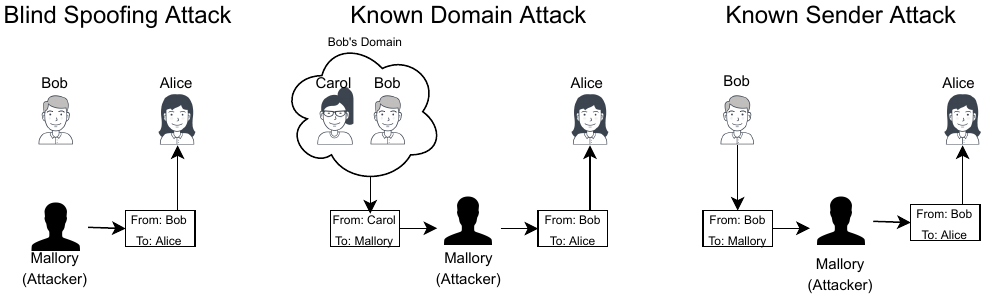}
\caption{The three simulated attack methods borrowed from Gascon et al. \cite{gascon2018reading}. In Blind Spoofing the attacker crafts a spear phishing email without any external information about the structure of the emails of the impersonated sender or their email domain. In Known Domain the attacker has access to emails from other individuals from within the impersonated sender's email domain and uses this information to forge spear phishing attacks with traits unique to this domain. In Known Sender the attacker has access to previous emails from the impersonated sender and this information to forge spear phishing attacks with traits unique to the impersonated sender.}
\label{fig:attacktypes}
\end{figure}

We target three attacks scenarios similar to \cite{gascon2018reading}. These are (i) \textit{Blind Spoofing}, (ii) \textit{Known Domain}, and (iii) \textit{Known Sender}. These attacks represent different scenarios in which the attacker has different levels of information about the impersonated sender. \textbf{(i)} The simplest of these attacks, \textit{Blind Spoofing}, attempts to simulate a scenario in which the attacker has very little information about the sender they are claiming to be. In this scenario we simply take a legitimate email and forge the sender address to that of a different sender. This aims to simulate a scenario in which the attacker doesn't have any information about the sender they are trying to impersonate beyond the sender's email.  While blind spoofing is a common technique deployed in other email-based attacks and not unique to spear phishing, it is used here because it is a crucial component of a large amount of spear phishing attacks and has shown to be capable of bypassing various security protocols and authentication methods \cite{shen2021weak}. 

\textbf{(ii)} The second, more sophisticated attack, is a \textit{Known domain} attack. In this scenario, the attacker has access to emails that belong to different senders within the same domain as the sender they are trying to impersonate. Therefore, the attacker will be able to forge transport features that are common between senders within the same domain. To simulate this attack, we take legitimate emails and change the sender address to a different address within the same domain. The reasoning behind this is that the emails of two different senders from within the same domain will have composition and transportation features that are the same but will still have behavioural features that are unique to the two senders. So by simply changing the email address to a different sender within the same domain, we can simulate a scenario where the domain-specific features that could have been used to successfully detect blind-spoofing attacks are no longer adequate for detecting this more advanced spear phishing attack. Essentially we limit the avenues that the classifier can take to identify spear phishing attacks and see how they perform when there are less clues to work with.  

\textbf{(iii)} The final attack method, which is the hardest to detect, is \textit{the Known sender}. In this scenario the attacker has access to emails from the sender they are impersonating, allowing them to incorporate the sender's features into their crafted emails and accurately impersonate the sender. This is replicated by taking an email from the sender to be impersonated, and changing the intended recipient. This is done under the assumption that with access to prior emails sent by the impersonated sender, the attacker would be able to forge all previous domain-specific features as well as the behavioural features that are unique to the sender with very little difference between the crafted spear phishing email and a legitimate email from the impersonated sender. As such this kind of attack is very difficult to detect and is intended to push the classifiers to their limits.

\begin{figure*}[ht]
\includegraphics[width=1\linewidth,scale=1]{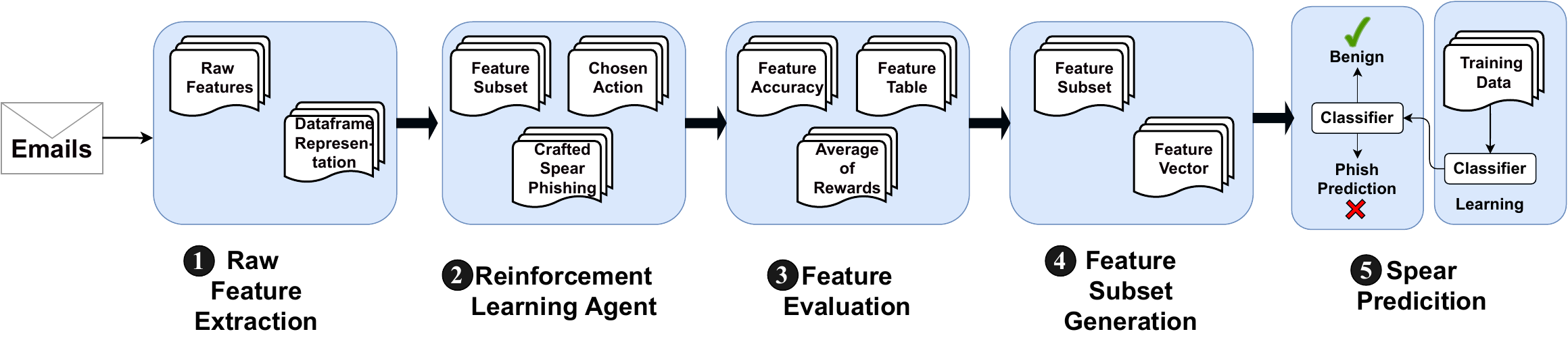}
\caption{Overview of RAIDER. Emails are taken as input. From here a feature subset is generated based on what features get the highest accuracy when detecting spear phishing attacks. From here the final feature subset is used for future predictions.}
\label{fig:RAIDER_flowchart}
\end{figure*}
\subsection{Overview of RAIDER}

We developed RAIDER, a system that automatically selects the best features for detecting spear phishing emails, and predicts whether an email is a spear phishing attack or a benign email. Figure \ref{fig:RAIDER_flowchart} depicts the workflow of RAIDER. In the following, we explain the steps in detail. A detailed overview of the reinforcement learning process is presented in figure \ref{fig:attacktypes}. This figure walks through a single step within the reinforcement learning process and shows how the reward for an action \& state pair is calculated. It also shows how these values are stored in and retrieved from RAIDER's AOR table.

\subsection{RAIDER in detail}\label{RAID_detail}

\subheading{\ballnumber{1} Raw Feature Extraction.} The first phase in RAIDER is the feature subset generation. In this phase, each field within an email is extracted and considered a `raw' feature. The raw features are then evaluated for their importance in detecting spear phishing attacks using our RL-based system. The features most useful for detecting spear phising attacks are those that can uniquely identify the behaviour of individual senders. Composition and transportation features that are unique to different email domains and email clients can also help flag incoming emails are spear phising attacks. While RAIDER does not directly interact with the body of the email, and therefore does not directly interact with email attachments, by using information and metadata from the header it can learn sender behavioural characteristics and detect spear-phishing attacks from non-text based features. 

When a representation of the email dataset is produced, the features are organized in a vector form according to the bag-of-words model\footnote{https://artsandculture.google.com/entity/bag-of-words-model},  where each email is represented as a matrix of integers, each integer denoting the frequency at which a specific word from within the dataset appears within that email. The matrix contains integers for every word within the dataset. The bag-of-words approach is used simply as a way to represent the header data of an email and determine the existence of certain traits within the header. This results in the aforementioned array which represents the existence and frequency of certain traits within an email.    

\subheading{\ballnumber{2} RL Agent.}
Choosing the specific feature to evaluate is decided by our RL agent. The RL agent chooses an action either by getting the best possible action from the feature table, or by randomly choosing an action from the action space. This process of selecting a feature, adding it to the feature subset, determining the feature effect, and updating the feature table represents a single step within the RL environment. After a step is performed, the resulting state is returned to the agent. In RAIDER the state is the current feature subset and corresponding accuracy. After a specified number of steps, the round finishes. After each round, the feature subset up to that point is discarded and the RL agent starts from scratch; allowing us to evaluate the importance of features within a variety of circumstances. A feature may only result in favorable/unfavorable accuracy changes when paired with other features. Therefore, it is important to evaluate features within a variety of circumstances to get a better understanding of the importance of each feature independently. The feature table is not reset and continues to be updated throughout the rounds.

\subheading{\ballnumber{3} Feature Evaluation.} RL is used to evaluate feature importance adapting Average of Rewards (AOR) approach introduced in~\cite{fard2013using}. The RL agent's action space consists of adding different possible features. The state in the environment consists of the current feature subset and the corresponding accuracy of the selected subset. Given a state the agent attempts to add features that will result in the highest possible accuracy increase. \textit{The agent determines the importance of each feature by creating a variety of feature subsets, testing these subsets with KNN, and determining how much of the increase in accuracy each feature is responsible for.} 
 
In RAIDER, the Feature importance is determined by the AOR and the times a feature was added. 
This latter value is incremented each time a feature is chosen during the RL phase. It is used to calculate the AOR and when generating the feature subset after the RL phase has ended. The action space for the agent is as follows: the agent can choose to add any one of the raw features to the existing subset. The specific feature is chosen by either exploiting the feature table by choosing the feature with the highest AOR, or by exploring the action space by randomly selecting a feature.

The accuracy increase/decrease for a feature is calculated by making predictions with KNN using the current feature subset. After the RL agent has chosen a feature, 
new training and testing datasets are generated using the new feature subset to make predictions. The training dataset contains benign emails only and the sender's email address serves as the label for each data sample. In this way, KNN is learning what values and patterns in the data represent what senders. This is also to be consistent with the assumptions in \cite{gascon2018reading} which `sender profiles' are generated for each sender. The testing dataset is a set of emails, 50\% of these  are benign, whereas the remaining 50\% are spear phishing emails that we have crafted according to the different attack methods previously outlined. KNN, using the learnt sender profiles, attempts to predict what sender sent the email in question. If the predicted sender address is different from the address on the email, we assume that this is a spear-phishing attack—\textit{someone masquerading as the sender rather than the sender themselves}. The predictions are then compared to the real status of the test emails and an accuracy is calculated for that feature subset. Any changes in accuracy between steps are calculated and using this information the feature table is updated with the new AOR (assuming there was some change in accuracy generated by adding the feature.)

\subheading{\ballnumber{4} Feature Subset Generation.} After a specified number of rounds, a final feature subset is created based on the values in the feature table. Any features that produced a positive increase in accuracy will be added to the final feature subset, whereas any features that resulted in a decrease, or no change, in accuracy will be removed. Any features that were not called during the RL phase will also be removed. This process allows us to generate a set of features to identify spear phishing attacks without any manual feature engineering. The process is fully automated by simply determining how each feature affects accuracy. This method also allows us to generate the features best suited for different attacks, and can adapt to zero-day attacks. A new feature subset can simply be generated as new threats emerge without the need for manual feature evaluation and engineering.

\begin{figure}[h!]
\centering
\includegraphics[width=0.5\linewidth,scale=1]{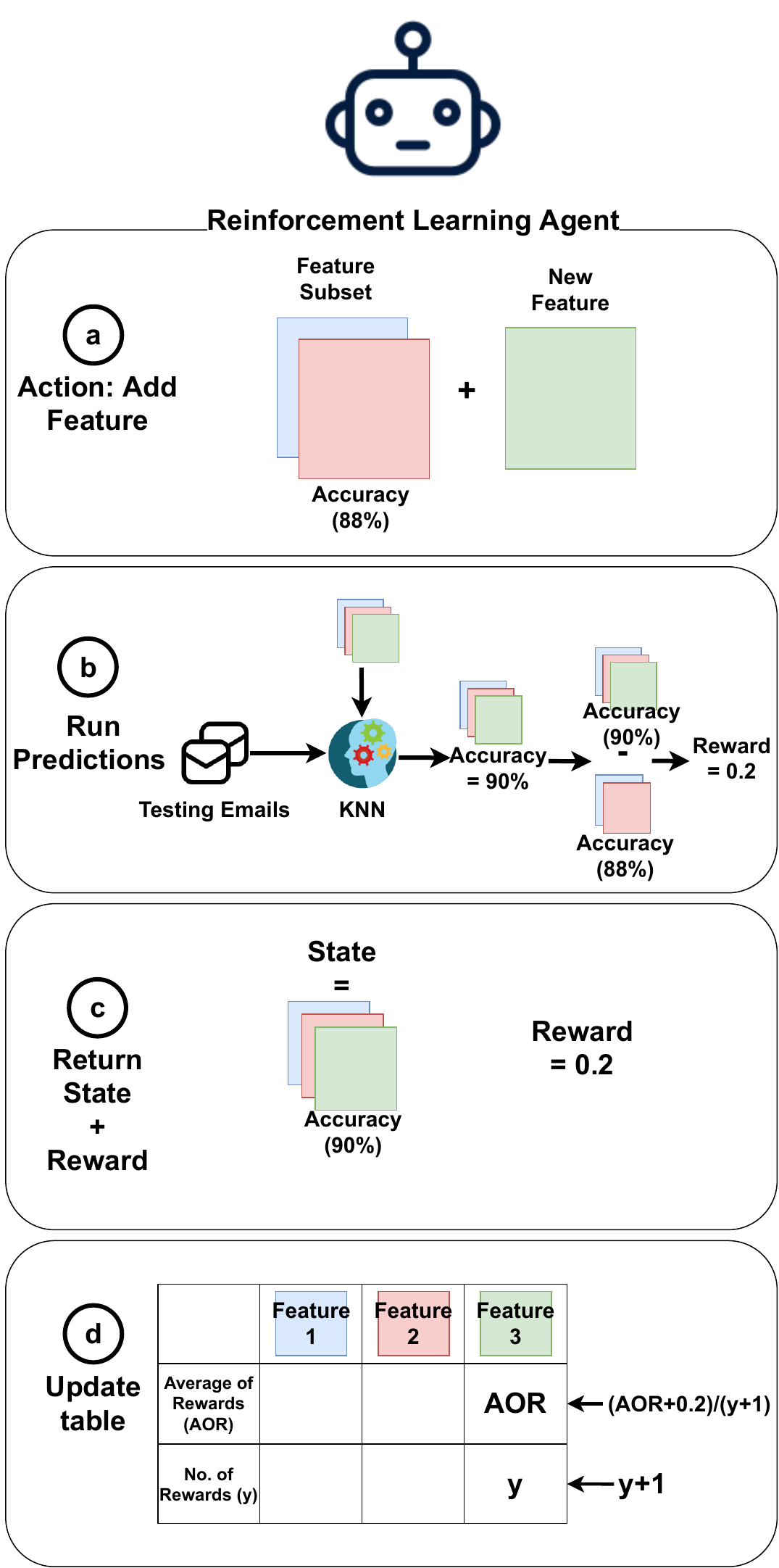}
\caption{Overview of RAIDER's reinforcement learning process that is used to evaluate raw features and generate feature subsets.}
\label{fig:attacktypes}
\end{figure}

\subheading{\ballnumber{5} Spear Prediction.} After the RL process is complete and a feature subset has been generated, RAIDER can then make predictions on incoming emails. At this point the algorithm is no longer being trained, and simply makes predictions on whether an incoming email is spear phishing or not. These emails are represented according to the feature subset. So for every incoming email a feature vector is generated that represents each email using only the features specified during the RL phase. RAIDER can be retrained and new features produced automatically if there are changes within the incoming data and the initial model can no longer adequately detect spear phishing emails.

%% file: Sections/5_Evaluation.tex
\begin{table*}[t]
\centering
\caption{Datasets used in the evaluation of RAIDER. During training sender profiles were built using emails in these datasets, with the exception of the CSIRO dataset where it is independently used as a measure for feature dimensions analysis. During testing spear phishing emails were crafted based on information obtained from these emails.}\label{tb:datasets}
\begin{tabular}{l|l|c}
\hline
\multicolumn{1}{c|}{Dataset} & \multicolumn{1}{c|}{Description}                                                                                                              & Emails \#\\ \hline\hline
Enron Corpus                  & Collection of emails of Enron Corporation                                                                          & 4,279          \\ \hline
Uni. Buffalo Dataset & Bread Secured program from University at Buffalo                                                                             & 75            \\ \hline
SpamAssassin easy ham         & Legit emails from Apache’s SpamAssasin                                                                                        & 2,551          \\ \hline
SpamAssassin hard ham         & Legit emails from Apache’s SpamAssasin                                                                                        & 250           \\ \hline
IWSPA-AP                      & \begin{tabular}[c]{@{}l@{}}Emails from IWSPA 2018\end{tabular} & 4,082          \\ \hline
CSIRO Dataset & Phishing emails(2016 to 2020) from CSIRO & 32,959
\\ \hline
\end{tabular}
\end{table*}

\section{Evaluation}
This section describes the experimental setup and performance evaluation for RAIDER.
\subsection{Experimental Setup}
We introduce our datasets, two phases of experimental settings (Reinforcement Learning and Zero-Day Simulation), and evaluation metrics below.

\subheading{Datasets.}
For our experiments we used a combination of publicly available email datasets (Table \ref{tb:datasets}). These datasets were chosen in order to have a diverse selection of emails to test our algorithm on. In order to avoid bias and over-fitting our system to any one dataset or email type we sourced a variety of different emails from different sources. The datasets used are as follows: The Enron Corpus, a collection of emails from the Enron Corporation prior to the company's collapse in 2001~\cite{ocampoh_2017}, University at Buffalo's Bread Secured dataset~\cite{bread}, Apache's SpamAssasin `easy ham' and `hard ham' datasets~\cite{apache}, and a dataset of political emails from the International Workshop on Security and Privacy Analytics (IWSPA-AP) 2018~\cite{balakrishnan_2018}. All of these email datasets are assumed to be comprised solely of benign emails sent from legitimate senders.

As previously mentioned, the spear phishing emails that we used for our experiments were crafted by us and used publicly available benign emails as a source of information when imitating senders. Spear phishing emails of varying sophistication were crafted according to the attack methods in section \ref{tmodel}. The sender address of the emails as well as domain and sender-specific traits were all taken from these benign datasets. Every feature is extracted from the email header and sensitive information such as the body of the email is left untouched. Our reasoning behind this approach is that the headers of emails contain not only information that is unique to each sender, but also information unique to email domains. We operate under the assumption that (1) the majority of spear phishing attacks are not at the level of complexity and sophistication where all uniquely identifying fields can be accurately forged, and (2) that even in sophisticated attacks where the attacker has access to significant information to create their deception, it is still possible for holes to exist within an attacker's masquerade and for certain uniquely identifying fields to be left unaltered. Therefore we use RAIDER to attempt to identify which fields within an email's header provide the highest level of reliability in uniquely identifying different senders and detecting different attacks. In the three different attack scenarios, the spear phishing emails are crafted according to the level of information the attacker is assumed to have about the sender. The more information the sender has, the more closely the spear phishing email resembles a benign email of the same impersonated sender. 

Essentially we attempted to impersonate the senders in these public datasets with our crafted spear phishing emails. The crafted spear phishing emails are consistent with \cite{gascon2018reading}. We incorporate email forgery techniques and take into account different scenarios based on how much information the attacker has on their target. 

When training KNN, whether it be within an iteration of the reinforcement learning process or for our prediction experiments, 7518 benign emails are used to train KNN and learn the sender classes. We imposed the rule that there must be at least two emails belonging to each sender within the training data. This is because one email isn't enough to identify traits common to emails sent by the same user. Ultimately this reduces the final number of benign emails to 8719 and results in 987 unique senders. During the testing phase 1201 benign emails and 1201 crafted spear phishing emails are used for making predictions, therefore giving the testing dataset a 50/50 split between benign and spear phishing emails. These 1201 benign emails are different from those used during training and are used to see whether the algorithm can correctly identify benign emails it hasn't seen before. Therefore of the total 8719 benign emails, 80\% of these are used for training and the remaining 20\% are used for testing. The 1201 spear phishing emails are crafted according to the attack method being used for the experiment and are used to see whether the algorithm can correctly identify spear phishing attacks.   


\subheading{Experimental Settings.}
RAIDER utilises an off-policy algorithm in which the behaviour policy is the Epsilon-Greedy policy. The behaviour policy is followed by the agent while the target policy is improved. This allows us to sufficiently explore all of our large numbers of features. Exploiting previous values too much would result in our agent neglecting the yet unexplored features.

To simulate a zero-day attack, the RL algorithm is trained on one attack type and then during the testing phase one of the other previously unseen attack types is introduced. We compared accuracy between static training in which the feature subset is never updated, and online learning in which the subset is generated in response to new attacks.

\subheading{Evaluation Metrics.}
To evaluate the results of our experiments the primary metric we make use of is accuracy. Accuracy in regards to Spear Phishing classification refers to the proportion of emails that were correctly predicted as either spear phishing or benign. Accuracy is defined as: \begin{equation}
    Accuracy = (TP+TN)/(TP+FP+FN+TN)
\label{eq:accuracy}
\end{equation}
where P/N (Positives/Negatives) refers to predicted outcome and T/F (True/False) refers to the actual outcome. For instance, TP (True Positives) refers to the proportion of emails predicted by the classifier to be spear-phishing that have been correctly identified.

\subsection{Results of RAIDER}

\subheading{Effectiveness of RAIDER.}

Table \ref{tb:accuracy} shows that by using automatically generated features RAIDER is able to detect spear phishing emails with slightly better or comparable accuracy to the state of the art \cite{gascon2018reading}. Obtaining equal (and in the case of blind spoofing, superior) results while eliminating the need to manually engineer features results in considerable time and effort saved. Data preparation such as cleaning data and engineering features accounts for 80\% of the work for data scientists and 76\% of data scientists view this as the least enjoyable part of their work \cite{press_2016}. It has also been demonstrated that different classifiers produce different results with the same set of features \cite{heaton2016empirical}, so in order to maximise the efficacy of a classifier, features will have to be engineered specially for that classifier and not reused. Therefore, being able to automatically generate features saves a lot of time and effort when engineering lots of features. Manually engineering features also limit the transferability of the feature set as features are built according to the problem being solved and can not be applied to other use cases. 

\begin{figure}[ht!]
\centering
\includegraphics[width=0.8\linewidth,scale=1]{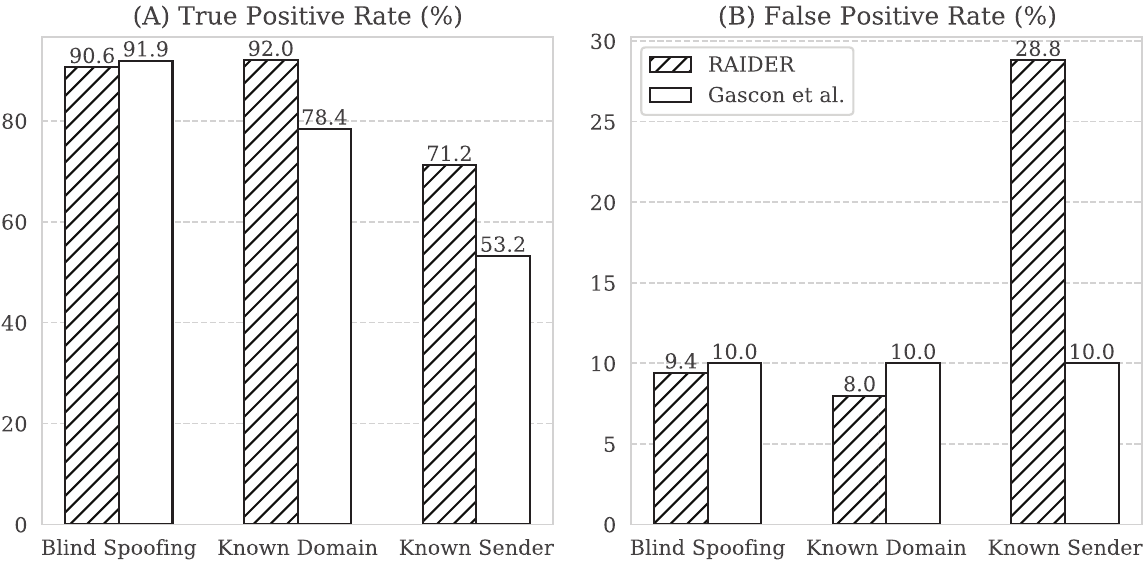}
\caption{Comparison of (A) TP rate and (B) FP rate of RAIDER with \cite{gascon2018reading}.} 
\label{fig:tpr_fpr}
\end{figure}

We compared the true positive rate and false positive rate of RAIDER with Gascon et al.'s \cite{gascon2018reading} KNN implementation when detecting spear phishing emails. This is depicted in Figure \ref{fig:tpr_fpr}, where the True and False Positive (TP/FP) rates of both systems are presented alongside one another. It is obvious that RAIDER performs better in detecting  TP. RAIDER also has less FP in the two more realistic threat models named \textit{Blind Spoofing} and \textit{Known Domain}, but performs worse than the state of art when it comes to the hardest yet rare threat model named \textit{Known Sender}.

\begin{figure}[ht]
\label{fig:prec}
\centering
\includegraphics[width=0.8\linewidth,scale=1]{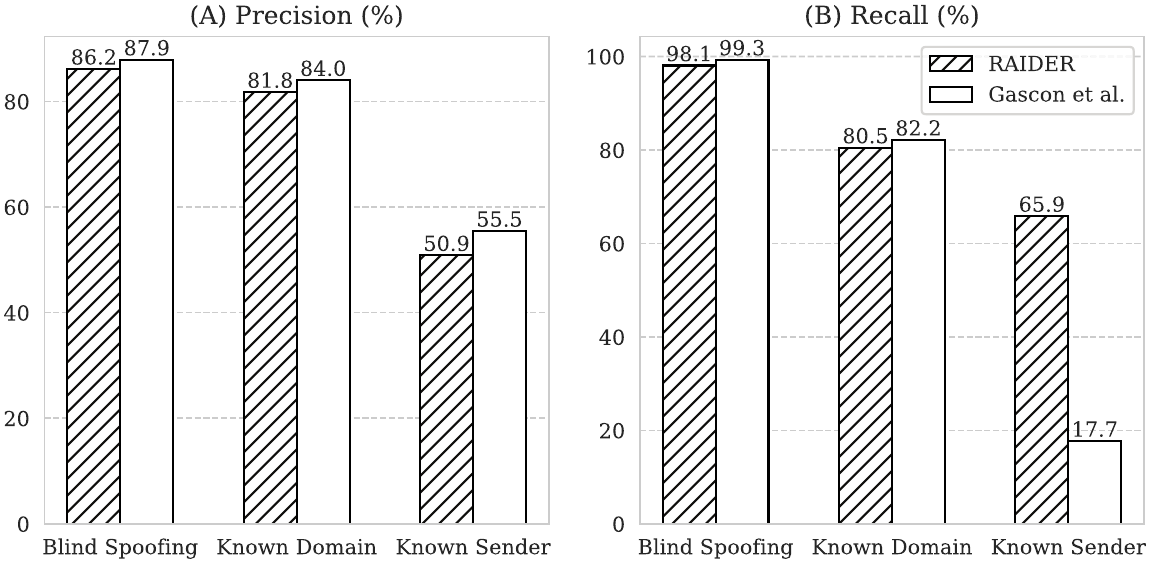}
\caption{Comparison of (A) Precision (B) Recall of RAIDER with \cite{gascon2018reading}.} 
\end{figure}

Figure \ref{fig:prec} shows that RAIDER and the manual KNN implementation obtain comparable precision and recall, with the manually extracted features just beating the automatically extracted features. Across all three attacks the manually-extracted features slightly beat the automated feature extraction when it comes to precision. I.e. of all the emails labeled as spear phishing more of these classifications were correct when using the manually extracted features than automatic. The recall shows that for all three attacks a higher percentage of the total spear phishing attacks were detected using the manually extracted features with the exception of the known sender attack. For the known sender attack, more than half of the spear phishing emails were correctly identified by RAIDER's automatically extracted features, whereas less than 20\% were correctly identified by KNN using the manually-engineered features. While the overall accuracy is about equal between the two systems, this suggests that RAIDER's automatically generated features are better identifying known sender spear phishing attacks whereas the manually engineered features more commonly identify spear phishing emails as benign.

\begin{figure}[ht]
\label{fig:roc}
\centering
\includegraphics[width=1\linewidth,scale=1]{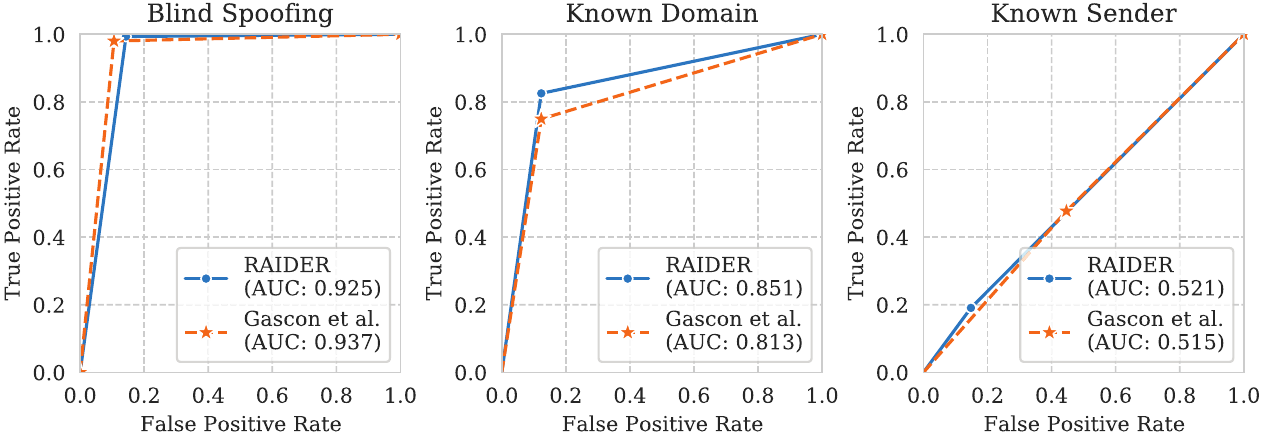}
\caption{ROC curve of RAIDER and Gascon et al. \cite{gascon2018reading}.} 
\end{figure}

Figure \ref{fig:roc} shows the ROC curve for RAIDER's automatically generated features and the manual KNN implementation against the three different attack types. The results suggest that there is very little difference in the performance of the two feature sets with RAIDER performing slightly better in two of the three attacks. 

The evaluation metrics performed suggest that the two feature subsets have comparable performance in terms of accurately classifying spear phishing emails. This shows that we are able to obtain classification accuracy comparable with state-of-the-art systems while automatically extracting features and eliminating the need for the manual feature engineering process. This not only saves time, but allows us to eliminate possible manual errors.

\begin{table}[t]
\centering
\caption{Comparison of accuracy results for RAIDER and KNN with manually engineered features across a variety of attacks. Percentages represent the number of emails RAIDER correctly predicts as either spear phishing or benign.}\label{tb:accuracy}
\begin{tabular}{c|c|c}
\hline
\begin{tabular}[c]{@{}c@{}}Attack \\ Scenario\end{tabular} & \begin{tabular}[c]{@{}c@{}}RAIDER \\ (Automatic feature subset)\end{tabular} & \begin{tabular}[c]{@{}c@{}}KNN \\ (Manually  engineered\\ features)\end{tabular} \\ \hline
Blind Spoofing                                             & 94\%                                                                                   & 90\%                                                                              \\ \hline
Known Domain                                               & 83\%                                                                                   & 83\%                                                                              \\ \hline
Known  Sender                                              & 62\%                                                                                   & 62\%                                                                              \\ \hline
\end{tabular}

\end{table}

\subheading{Robustness of RAIDER.}

\begin{figure}[t]
\centering
\includegraphics[width=0.7\linewidth,scale=1]{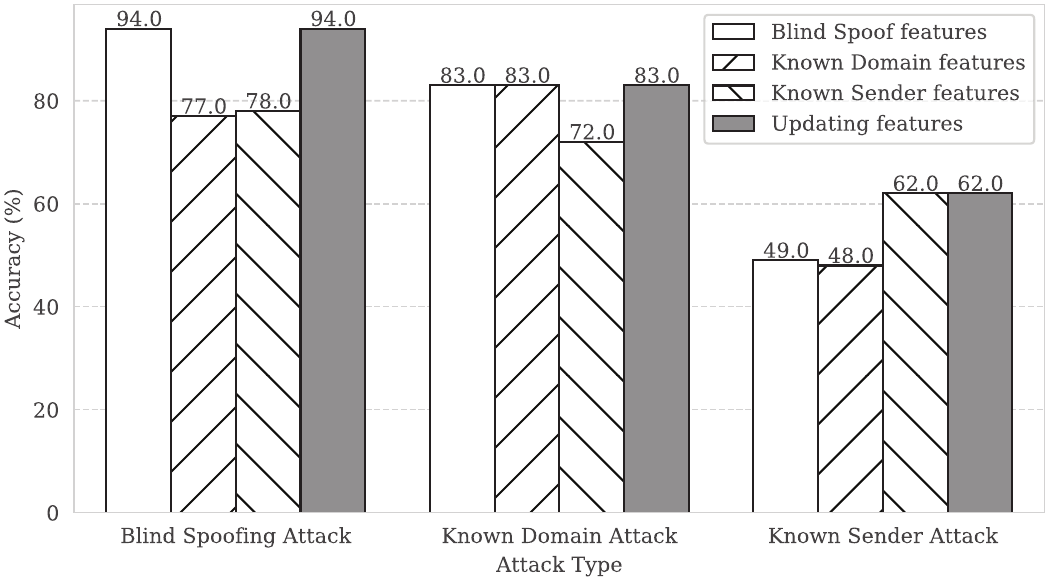}
\caption{Accuracy of the different feature vectors over time. Feature subsets are generated based on one kind of attack, then are tested with crafted spear phishing emails from all the different attack types. Graph shows how the accuracy of predictions changes when new attacks are introduced. Accuracy refers to the percentage of emails that are correctly identified as either benign or spear phishing. The x axis describes the attack type used for the testing of RAIDER, with the accuracy on the y axis showing the accuracy of each variation of RAIDER against the type of attack on the x axis. Each bar represents a different variation of raider where the independent variable is the type of training data used. Training data can either be blind spoofing attacks, known domain, known sender, or updating features. For any of the first 3 the features were derived using only spear phishing emails using that type of attack. So for blind spoofing, a feature subset was generated using only blind spoofing spear phishing emails. For updating features, the feature subset is updated every time a new type of attack occurs. So when known domain attacks are introduced to RAIDER, it then retrains using known domain attacks and produces a new feature subset. This attempts to simulate a scenario where RAIDER updates the feature subset in response to new attacks.}
\label{fig:0day_graph}
\end{figure}

Figure \ref{fig:0day_graph} indicates that for any given attack method (with the exception of known domain attack), the feature subset that was generated using attacks of the same type always returns the highest level of accuracy. This trend is most significant in the case of blind spoofing and known domain attacks. The `Updating features' column shows an implementation of RAIDER where a new feature subset is generated every time a new attack type is encountered. So if the system was originally trained using blind spoofing attacks, when known domain attacks appear RAIDER will retrain using known domain attacks and produce a new feature subset. The 3 other columns train using only one type of attack and never update regardless of what attacks they encounter. The updating features column consistently obtains the highest accuracy whereas the systems that don't retrain experience lower accuracy when detecting attacks they have not previously seen.  

The blind spoofing subset returned an accuracy rate of 94\% when predicting blind spoofing spear phishing emails, while the known sender subset returned a rate of 78\% and known domain returned 77\%  —A difference of 16\% and 17\% respectively. Testing known sender attacks with the known sender subset returned an accuracy rate of 62\%, an increase of 14\% over the known domain subset, and 13\% over the blind spoofing subset. We believe this demonstrates our system's ability to adapt to different attack methods and previously unseen threats by generating a new feature subset in response to changes within the data. In this scenario, the subset from one kind of attack being applied to another kind of attack represents a prediction system that has been previously trained statically at one point in time and is now encountering previously unseen data. Therefore, we believe that by leveraging RAIDER's ability to automatically generate feature subsets, the system can be updated to better detect new types of attacks. Although it is also worth noting that while this method provides higher accuracy than statically trained models, there is still a decrease in accuracy between attack types, regardless of whether the feature subset is generated or not. This is expected as each attack has a varying level of sophistication and complexity.

Our experiments thus far have simply compared the prediction abilities of a statically-trained implementation of RAIDER to that of a dynamic one that updates the feature set to adjust to new attacks. To comprehensively demonstrate RAIDER's zero-day capabilities further testing may need to be done to compare state-of-the-art statically trained systems to RAIDER, to see how accuracy is affected by the emergence of new threats. However, the detailed 3 attack methods we picked are not suitable for testing the ability of Gascon et al's system \cite{gascon2018reading} to detect zero-day attacks, as all of their manually engineered features are intended to be used across all 3 of these attack types. Therefore, to sufficiently test the state-of-the-art manual feature engineering system, we would need to craft a variety of different attacks that the current feature set is not based around which we found to be very challenging to achieve in practice.

\begin{figure}[t]
\centering
\includegraphics[width=0.8\linewidth,scale=1]{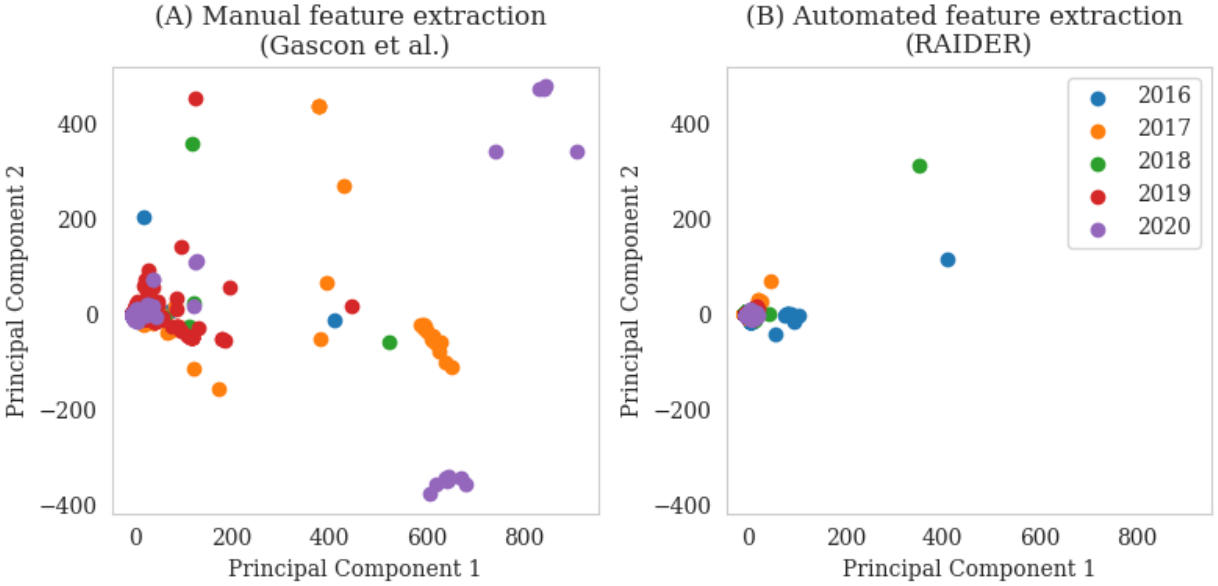}
\caption{Principal Component Analysis (PCA) scatterplot of the two feature engineering methods. Here PCA projects the data points into a new 2D space while the axes do not have physical meaning. The new values on x-axis and y-axis (Principal Component 1 and 2) contribute most to the variation through a transformation.}
\label{fig:dimension1}
\end{figure}

\subheading{Feature Stability.}
We then conducted Principal Component Analysis (PCA) on our automated feature extraction process, as well as the manual process based on Gascon et al.'s work \cite{gascon2018reading}. We did this using CSIRO's phishing email dataset which contains over $32,000$ emails from 2016 to 2020. 
As shown in Figure \ref{fig:dimension1}, we reduce the dimension and visualise the features into a new 2D space. The features generated by our automated process shows  stronger stability and less dimensionality over the years compared to \cite{gascon2018reading}.

\begin{figure}[thb]
\centering
\includegraphics[width=0.8\linewidth,scale=1]{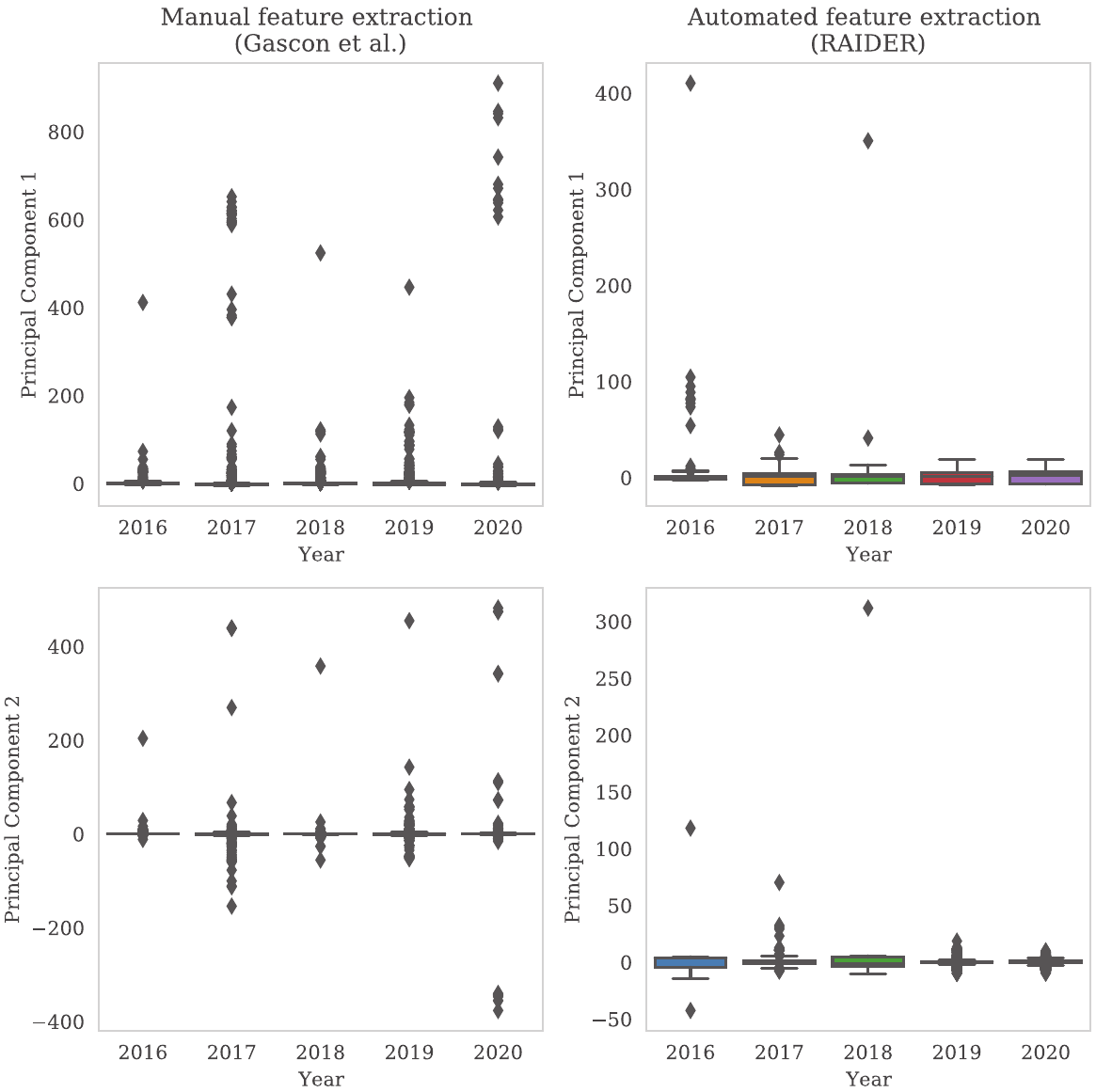}
\caption{Boxplot representation of PCA.}
\label{fig:dimension2}
\end{figure}

RAIDER is able to gain this comparable accuracy with feature vectors of significantly smaller dimensions than those of state-of-the-art spear phishing feature extraction. Figures \ref{fig:dimension1} and \ref{fig:dimension2} show that excluding a few outliers, emails extracted using RAIDER remain clustered with low dimensionality, whereas those extracted using Gascon et al.'s features are spread out more, meaning that their feature vectors are more volatile and more frequently reach higher dimensions. This means that our system manages to achieve comparable, and in certain cases superior, accuracy to state-of-the-art spear phishing detection systems while maintaining smaller feature vectors. Looking at 
Figure \ref{fig:dimension1} shows that our feature vectors are more tightly concentrated while also have a smaller upper-bound than state-of-the-art feature extraction.

%% file: Sections/6_Discussion.tex
\section{Discussion and Future Work}

\subsection{Complexity and Time overhead}
The runtime cost was recorded for making predictions with our manual feature extraction process as well as RAIDER's automated feature extraction. For both methods, we recorded the time it takes to pre-process and prepare the datasets, as well as the time taken to train and perform predictions using KNN. Additionally, we also recorded the time taken by RAIDER to evaluate and extract features using the reinforcement learning process. 
Due to the difficulty of quantifying the time requirements of the manual feature engineering process, it is difficult to make a meaningful comparison between manual and automated feature engineering. The automated feature extraction process is more computationally demanding and results in a longer runtime but circumvents the prolonged real-world time and labour requirements of manual feature engineering. It also alleviates the risk of human error.

\begin{table}[t]
\centering
\caption{The runtimes for both the system using manually engineered features and the system using automated feature extraction. All times are in seconds.}\label{tb:time}
\begin{tabular}{l|c|c|}
\cline{2-3}
\multicolumn{1}{c|}{\textbf{}} & \textbf{\begin{tabular}[c]{@{}c@{}}KNN (Manual \\ features)\end{tabular}} & \textbf{\begin{tabular}[c]{@{}c@{}}RAIDER (Automated \\ feature extraction)\end{tabular}} \\ \hline
\multicolumn{1}{|l|}{\textbf{\begin{tabular}[c]{@{}l@{}}Dataset preprocessing\\ time\end{tabular}}}  & 136.84          & 317.8            \\ \hline
\multicolumn{1}{|l|}{\textbf{\begin{tabular}[c]{@{}l@{}}Reinforcement\\ learning time\end{tabular}}} & n/a             & 1489.81          \\ \hline
\multicolumn{1}{|l|}{\textbf{Prediction time}}                                                       & 23.93           & 20.65            \\ \hline
\multicolumn{1}{|l|}{\textbf{Total (Seconds)}}                                                       & \textbf{160.77} & \textbf{1828.26} \\ \hline
\end{tabular}
\end{table}

The results in table \ref{tb:time} show that RAIDER's automated feature extraction process is more computationally complex and results in a longer run-time than using the manually engineered features. This is due to two reasons. The first is that the pre-processing and representing of the dataset takes longer at 317.8 seconds for the automated feature extraction process and 136.84 for manually extracted features. This is due to the fact that because RAIDER initially extracts every field within the email header before the reinforcement learning phase it is working with significantly more features than the 32 manually extracted features. The second reason being that the reinforcement learning process takes considerable time (1489.91 seconds) to accurately determine the importance of each feature. Ultimately in a real-world scenario, the reinforcement learning process would not be executed very frequently and while there is more computational overhead associated with automated feature extraction, there is considerable time and human labour costs avoided by eliminating the need for manual feature engineering. The automated feature extraction process can also adapt to new attacks and evolve whereas the system that uses manually engineered features cannot.

In addition, after the initial deployment and fitting of the algorithms, the actual time cost of classifying emails is slightly better for automated features. Using the manual features resulted in a prediction time of 23.93 seconds for all 2402 testing emails, or 0.009962531 seconds/email whereas using the automated features resulted in 20.65 seconds overall or 0.008597002 seconds/email.    

\subsection{Contextual Analysis Limitations}
In its current implementation RAIDER analysis and classifies each unique email independently. This means that if an email being analysed by RAIDER is a part of a conversation in which previous emails have already been sent between the sender and the victim, RAIDER would neglect the previous correspondence and make predictions using only the single email that is currently being analysed. I.e. it would attempt to classify the email based only on that email's similarity to the training data, and would not take into consideration its context within the ongoing conversation between the sender and the victim.

This could potentially limit the efficacy of RAIDER as the existence of a spear phishing attack may only be apparent using information sourced from multiple emails. It is also possible that if an attacker sends multiple emails, some of these may be flagged as spear phishing while others avoid detection. Obviously, if one email within a chain is a spear phishing email, the legitimacy of the other emails sent by this sender should be called into question and the sender be considered untrustworthy.  

This could also improve the accuracy at which the more sophisticated known sender attacks are detected. By taking into consideration entire conversations rather than just individual emails it may be more difficult for attackers to disguise themselves as a legitimate entity as they would require more information. Scenarios in which the attacker has access to a single email from the sender (Known sender attacks) may no longer deal such a critical blow to RAIDER, as behavioural characteristics that only become apparent across conversations may be difficult to replicate with limited information, thus increasing the information required by the attacker to create a deception possible of fooling RAIDER.

Possible further work for RAIDER involves expanding on this functionality and seeing how the efficacy of RAIDER is affected by a more contextually aware classification algorithm. At the very least a system to flag senders who have previously sent phishing attacks as untrustworthy could help prevent false negatives. 

\subsection{Crafted vs Real Spear Phishing Emails}
As previously mentioned, due to the targeted nature of spear phishing attacks, spear phishing emails based on publicly available benign emails were crafted for these experiments. This was due to a lack of availability of spear phishing datasets suitable for this research. Because the algorithms work by constructing sender profiles for each sender within a mailbox, if genuine spear phishing emails were to be used, multiple legitimate emails for a sender would be needed (both for training and testing), as well as genuine spear phishing attacks targeted towards that sender (for testing.) This kind of data is not readily available in large enough quantities for it to adequately test the systems used. 

The three different attack types are crafted based on real spear phishing techniques which have been proven to be frequently used in genuine spear phishing emails \cite{hu2018end,mohamed2020predictive,duman2016emailprofiler}. Despite their basis in real spear phishing attacks, they are still emails crafted for the purpose of testing spear phishing classifiers, and may fall short of fully replicating real-world spear phishing attacks and the phenomena of attacks evolving over time. As such, to fully understand the applicability of RAIDER, it may require real-world testing in which it can encounter genuine spear phishing emails. While we believe the research we have done addresses the computational requirements of existing spear phishing detection systems and proposes a time-proof system that can detect zero-day attacks, we acknowledge that there is still more research to be done to fully quantify the strength and weaknesses of our system and that while the data so far is promising, we can not categorically assume that the exact same results will be replicable in a real-world scenario. 

Ultimately the aim of RAIDER was to demonstrate the efficacy and practicality of using automatically generated features as opposed to manually engineered features. We believe we have achieved this and that the work we've done shows the transferability of RAIDER's framework to different attack types. Therefore, in the way that RAIDER can be applied to our multiple different attacks we believe that it will also be applicable to real-world spear phishing attacks. RAIDER's use in a real-world environment is outside the scope of this research but to us is the next logical step in evaluating its capabilities.

%% file: Sections/7_Related_Work.tex
\section{Related Work}

Current literature regarding the identification of spear phishing utilises technical controls and software solutions such as email authentication \cite{xiujuan2019detecting}, black-listing IP addresses \cite{agazzi2020phishing}, and ML techniques \cite{han2016accurate}. ML-based approaches have proved an effective method of detecting spear phishing attacks, but as far as we are aware there has been no previous work on detecting zero-day attacks with these systems. The current landscape of ML-based spear phishing detection is summarised based on the information that different implementations make use of to make predictions.

\textbf{Stylometric Features.} Stylometric or linguistic analysis involves generating features that represent a sender's style of writing. The idea here is that, an email sent by an attacker will have subtle differences in their style of writing than the sender they are impersonating, and that the presence of these differences suggests a spear phishing attack. Dewan et al. \cite{dewan2014analyzing} and Stringhini and Thonnard \cite{stringhini2015ain} implement systems that analyse  the linguistic style and behavioural patterns of emails such as the number of paragraphs a sender typically uses or whether they use a space after punctuation. \cite{duman2016emailprofiler} performed similar research into using stylometric features from the subject and body of an email to detect spear phishing attacks. Their solution also considers different deployment options such as having a remote trusted server that users can query when they receive an incoming email.  

\subheading{Email Header Metadata.} Previous research has also been done into analysing email header metadata to detect spear phishing attacks. Gascon et al. \cite{gascon2018reading} leveraged a variety of email headers as means to identify spear phishing attacks. They found that composition features such as the encoding of an email and transportation features such as the timezone path an email takes to reach its destination provide a means of validating an email when other fields have been spoofed. Bhadane and Mane \cite{bhadane2018detecting} made use of email metadata within an organisational setting, looking at scenarios where a spear phishing attack is being launched from a compromised legitimate account within a network. They made use of information such as IP addresses and an email's travel route to detect spear phishing attacks within a real-world scenario. Samad and Gani \cite{samad2020analyzing} used the metadata of email attachments to detect spear phishing emails.

\subheading{Misc.}
In addition to extracting information from the email in order to predict spear phishing attacks, there also exist studies that make use of external information sources when evaluating incoming emails. In addition to stylometric features, Dewanet al. \cite{dewan2014analyzing} also made use of information sourced from senders' LinkedIn profiles to determine if an email was spear phishing or not. Although they found that using these social features did not provide any benefit to prediction accuracy. Other studies \cite{das2019sok,bhadane2018detecting,ho2017detecting} queried \textit{WHOIS} and \textit{FQDN} information based on data retrieved from the email headers—usually combining these with stylometric and metadata features.

The work of Gascon et al. \cite{gascon2018reading} can be considered the best state-of-the-art work due to their balancing of high true positive rates with low false positives rates, and their privacy-friendly feature extraction method. As far as we are aware there has been no previous spear phishing detection approach that (a) selects significant features autonomously  in response to emerging threats, and (b) considers the scalability of the solution based on the existing required large feature vectors. Therefore, the aim of our work is to address those two challenges to produce smaller feature vectors that autonomously change in response to emerging threats.

\subheading{Feature Selection}
We identified the shortcomings of contemporary spear phishing detection systems as a result of the lack of optimization in the feature selection process. While to the best of our knowledge there has been no previous application of feature selection optimization within the realm of spear phishing, there has been considerable research into optimized feature selection. Stochastic optimization algorithms have been demonstrated as an effective way to find meaningful features within data with an extremely large number of possible variables \cite{gadat2007stochastic}. Dai \& Guo proposed a Beta Distribution-based Cross-Entropy framework capable of effectively selecting features across a variety of high-dimensional datasets \cite{dai2019beta}. Yamada et al. applied stochastic gates to non-linear classification and regression functions using a variety of real and artificial datasets \cite{yamada2020feature}. While stochastic algorithms have proven to be an effective method of optimizing the feature selection process, there have been no prior studies as to their effectiveness within the sphere of detecting spear phishing attacks. Although as the above works show certain stochastic frameworks have proven to be agnostic and transferable to different data types.

%% file: Sections/8_Conclusion.tex
\section{Conclusion}
In this paper, we have explored the possibility of using reinforcement learning to detect zero-day spear phishing attacks. We have devised a spear phishing email detection system (RAIDER) which uses reinforcement learning to automatically evaluate and select important features, while using KNN to make predictions on incoming emails. By simulating different spear phishing attack techniques of varying sophistication, we have demonstrated how our classifier responds when trained with different datasets. We have shown that our automatically generated feature sets (based on a reinforcement-learning algorithm) are of equal or better accuracy than systems, which use manually engineered features. We have also provided evidence that RAIDER saves time and effort in spear phishing identification tasks. The process of generating features takes, on average, 24.83 minutes and allows for a prediction accuracy of 94\% for blind spoofing attacks, which is one of the most widespread techniques employed by adversaries. We have also found that by automatically generating a feature subset, we can potentially detect previously unseen attacks and maintain a more consistent level of accuracy across different attack specimens than statically-trained alternatives. Our results show that by utilising the RAIDER's automatic feature generator, we can avoid accuracy drops of up to 14\% when encountering new (previously unseen) attacks.  

\section*{Acknowledgment}
The work has been supported by the Cyber Security Research Centre Limited whose activities are partially funded by the Australian Government’s Cooperative Research Centres Programme. This version of the contribution has been accepted for publication, after peer review (when applicable) but is not the Version of Record and does not reflect post-acceptance improvements, or any corrections. The Version of Record is available online at: \url{https://doi.org/10.1007/978-3-031-23020-2_2}. Use of this Accepted Version is subject to the publisher’s Accepted Manuscript terms of use \url{https://www.springernature.com/gp/open-research/policies/accepted-manuscript-terms}.